\documentclass[twocolumn,showpacs,preprintnumbers,amsmath,amssymb]{revtex4}
\usepackage{amsmath,amsthm}
\usepackage{graphicx}
\usepackage{dcolumn}
\usepackage{bm}

\begin{document}
\title{Gamow peak approximation near strong resonances} 
\author{Sachie Kimura}
\affiliation{Department of  Physics, University of Milano, via Celoria 16, 20133  Milano, Italy}
\author{Aldo Bonasera}
 \affiliation{Cyclotron Institute, Texas A\&M University, College Station TX 77843-3366, USA and \\
 INFN-LNS, via Santa Sofia, 62, 95123 Catania, Italy}


\date{\today}

\begin{abstract} 

We discuss the most effective energy range for charged particle induced reactions in a plasma environment   
at a given plasma temperature. 
The correspondence between the plasma temperature and the most effective energy should be modified from the one given by the Gamow peak energy, 
in the presence of a significant incident-energy dependence in the astrophysical $S$-factor
as in the case of resonant reactions. 
The suggested modification of the effective energy range is important not only in thermonuclear reactions at high temperature in the stellar environment, e.g., 
in advanced burning stages of massive stars and in explosive stellar environment, as it has been already claimed, 
but also in the application of the nuclear reactions driven by ultra-intense laser pulse irradiations. 

\end{abstract}

\maketitle

The nuclear reaction rate in a plasma environment at a certain temperature
can be related to an effective energy range~\cite{nacre,ueda,liolios,newton,rauscher2010}, both 
in the stellar site~\cite{nacre,daacv,newton} and in the laser-induced plasma site~\cite{ditmire, zweiback, buersgens,woosk, marina,belyaev:026406,abc}. 
This effective energy range gives us an idea of which energy region one can compare to the cross section data 
in the conventional beam-target experiments, when one needs to know the reaction rate in a plasma at a given temperature $T$.
Of particular interest is the application of this relation to the nuclear reaction yield in the laser-induced plasma site.
The neutron yield through the reaction $^2$H($d,n$)$^3$He driven by laser-pulse irradiation on 
deuterium cluster target is well studied~\cite{ditmire, zweiback, buersgens,woosk, marina} at various laser 
parameters. The deuteron acceleration in such an experiment is attributed to the Coulomb explosion of the clusters in the laser pulse field.  
By measuring the energies of accelerated deuterons,
 the laser-induced plasma deuterons are known to have Maxwellian-like energy spectra~\cite{buersgens}.    
Recent experiments on Texas Petawatt laser are dedicated for the determination of the deuteron plasma temperature~\cite{woosk, marina} 
by using cryogenically cooled deuterium D$_2$ (or near-room-temperature deuterated methane C$D_4$) cluster and $^3$He mixtures. By taking the ratio of the fusion yields from reactions $^2$H($d,n$)$^3$He, $^2$H($d,p$)$^3$H and $^3$He($d,p$)$^4$He, 
the temperature of the deuteron plasma is determined to be from 8 keV to 30 keV. 
The other examples are 
the proton induced reactions $^{11}$B($p,n$)$^{11}$C and $^{63}$Cu($p,n$)$^{63}$Zn using the laser-accelerated protons~\cite{spencer2} 
and the reaction $^{11}$B($p,\alpha$)$^8$Be.
For the former two reactions, protons are accelerated from a thin foil target and interact with the secondary solid targets. In such a case  
the mechanism of the ion acceleration is attributed to the target normal sheath acceleration~(TNSA) and 
the spectra of the laser-accelerated protons are, again, known to be a near-Maxwellian but with the temperature as high as 5 MeV~\cite{kb-sc}.   
For the latter reaction, 
the yield of 10$^3$ $\alpha$-particles has been reported by a group in Russia 
in the laser-pulse irradiation of  the peak intensity 2 $\times$ 10$^{18}$ W/cm$^2$ in 1.5~ps 
on the $^{11}$B+CH$_2$ composite target~\cite{belyaev:026406}. 
However no data is published on the spectra of the accelerated ions from this experiment and, besides, 
corrections taking into account the particles ranges in matter reveal a higher yield (10$^5 \alpha$)~\cite{PhysRevE.79.038401}.
The $\alpha$-particle yield through the reactions 
$^{11}$B($p,\alpha$)$^{8}$Be and $^{10}$B($p,\alpha$)$^7$Be are observed, using natural boron doped plastic (CH$_2$) targets~\cite{abc},  
at ABC laser facility, which derivers 50~J in 3 ns, in Frascati in Italy.
In this experiment the spectra of the accelerated ions of boron as well as protons are characterized, 
but the observed fusion yield is not 
fully consistent with the one expected from the characterized ion spectra.   
In the above mentioned experiments knowing the effective energy of the plasma ions which contribute to the nuclear 
reactions of interest is essential both 
to understand the acceleration mechanisms of energetic ions generated in the laser-plasma 
interaction
and for the optimization of the laser parameters using the scaling relation~\cite{kb-sc}. 
By comparing the ion spectra expected from the reaction yield 
 with the one obtained from the direct measurement 
of the accelerated ions, one can also determine the energy loss of the ions in the plasma~\cite{woosk}, which is not understood completely.
In this connection,  we mention that 
the recent experiments at TRIDENT laser at Los Alamos National Laboratory report a success of 
deuteron acceleration as high as 170 MeV from an ultra-thin (300 nm) foil target~\cite{prl2013} by the newly proposed break-out afterburner~(BOA) mechanism.   
The intense deuteron-beam generated by this mechanism is used to produce an intense neutron-beam by means of the reaction $^9$Be($d,n$)$^9$B. 
The promising result suggests a possibility of a compact neutron source generator driven by high-intensity laser pulses
and opens up various potential applications using the deuteron induced reactions which have an advantage of positive Q-values compared with proton induced reactions~\cite{kb-pet,koreanDeuteron2012}. 
We mention also that 
this relation between the plasma temperature and the effective energy can be applied the other way around~\cite{marina}. 
Through the measurement of fusion yields in a laser-induced plasma, one could determine low energy 
cross sections~\cite{prl_cs}, which are of great interest for astrophysical applications. 
For this purpose, one needs to know the exact relation between the plasma temperature and the most effective energy
not only in non-resonant reactions but also in resonant reactions.



When both colliding ion species have thermal distributions,
the reaction rate can be obtained by integrating 
reaction cross section~$\sigma$ multiplied by the relative velocity $v$ and by the 
spectrum $\phi(v)$ of the relative velocity over the incident energy $E$ (keV)~\cite{clayton,newton,rauscher2010,nevins,ueda}.
The relative velocity spectrum is written in a form of Maxwell-Boltzmann distribution 
in an equilibrium gas at temperature $T$. 
One, thus, obtains the reaction rate per pair of particles as a function of temperature: 
\begin{equation}
  \label{eq:sigmav}
  \langle\sigma v\rangle=\sqrt{\frac{8}{\mu \pi}}\frac{1}{(k_BT)^{3/2}}
  \int^{\infty}_0 S(E)\exp\left(-\frac{E}{k_BT}-\frac{b}{\sqrt{E}} \right),
\end{equation}
where $\mu$ is the reduced mass of the colliding nuclei 
and $b=31.28Z_1Z_2A^{1/2}$ keV$^{1/2}$, denoting the atomic numbers and reduced mass number of colliding nuclei 
$Z_1, Z_2$ and $A$, respectively.   
We write the cross section in terms of
the astrophysical $S$-factor, $S(E)$~\cite{newton,rauscher2010,PhysRevC.76.031602}; $k_B$ is the Boltzmann constant.    
The effective energy at a certain temperature $T$ is determined by means of the Gamow peak.  If 
$S(E)$ is a smooth function of energy, we may approximate $S(E)$ with a constant~($S_0$) and bring it 
out of the integral. Then, by means of the saddle-point approximation, 
we expect that the most important contribution to the integral in Eq.~(\ref{eq:sigmav}) comes from 
$E=E_0$ which satisfies the following condition:
\begin{equation}
  \label{eq:saddl}
  \frac{d}{dE}\left(-\frac{E}{k_BT}-\frac{b}{\sqrt{E}} \right)_{E=E_0}=0.
\end{equation}
This condition leads to the most effective energy $E_0$~(or Gamow energy) at temperature $T$,
\begin{equation}
  \label{eq:tgamow}
  k_BT=\frac{2}{b}E_0^{3/2}.
\end{equation}
The method of the saddle point approximation is equivalent to the replacement of the peak  by a gaussian with 
the same peak and with the $1/e$ width of 
\begin{equation}
  \label{eq:tdgamow}
  \Delta E_0=\frac{4}{\sqrt{3}}(E_0k_BT)^{1/2}.
\end{equation}   
The width is a function of the plasma temperature. 
In Ref.~\cite{nacre} the correspondence between $T_9$-axis and $E$-axis is derived from 
this equation. $T_9$ is the temperature in the unit of 10$^9$~K. $E_0 \pm \Delta E_0/2$ represents the effective energy window.

However
if the reaction cross section has a significant energy dependence,  
the astrophysical  $S$-factor cannot be approximated by a constant to evaluate the integral in Eq.~(\ref{eq:sigmav}).  
One, therefore, has to consider the contribution from this term in addition to 
the two terms in Eq.~(\ref{eq:saddl}). 
In such a case, practically, the 
condition to get the most effective energy  Eq.~(\ref{eq:saddl}) becomes,         
\begin{equation}
  \label{eq:saddl-2}
  \frac{d}{dE}\left(\log S(E)-\frac{E}{k_BT}-\frac{b}{\sqrt{E}} \right)_{E=E_0}=0.
\end{equation}
To discuss more concretely,  
 we consider an example of resonant reactions
where the astrophysical $S$-factor is approximated 
by the following Breit-Wigner form:
\begin{equation}
  S(E)=S_0+\frac{S_r}{(E-E_r)^2+\Gamma^2/4},
\end{equation}
where $E_r$ and $\Gamma$ are the peak and the width of the resonance. 
We assume that both $S_0$ and $S_r$ are positive.
Then Eq.~(\ref{eq:saddl-2}) leads 
\begin{widetext}
\begin{equation}
  \label{eq:kt}
  k_BT=\left(\left[\frac{-S_r 2(E-E_r)}{\left((E-E_r)^2+\Gamma^2/4\right)\left(S_r+S_0(E-E_r)^2+S_0\Gamma^2/4\right)}
       +\frac{b}{2}E^{-3/2} \right]_{E=E_0'} \right)^{-1}.
\end{equation}
\end{widetext}
By substituting $S_r=0$,  it can be, easily, derived that this equation recovers the conventional Gamow energy at 
temperature $T$ (Eq.~(\ref{eq:tgamow})). 
If $S_r$ is not zero, i.e., in the presence of a resonance, Eq.~(\ref{eq:kt}) implies that
 the departure of the most effective energy from the Gamow peak is 
large around the resonance peak $E=E_r$.  
Another simple limit is the case where the width of the resonance is zero and $S_0$ is negligible, i.e., 
the resonance is approximated by a $\delta$-function,  then
\begin{equation}
  \label{eq:ktd}
  k_BT=\left(\frac{-2}{E_0'-E_r}
       +\frac{b}{2}E_0'^{-3/2} \right)^{-1}.
\end{equation}
In this limit the absolute value of the correction term becomes larger than the second term and the resonant peak 
gives the major contribution. 
Given that both $S_0$ and $S_r$ are positive, 
the first term in this equation changes its sign as $E$ passes through $E_r$,
i.e., at a given temperature $T$ the most effective energy, $E_0'$, becomes higher than 
$E_0$ in the region $E<E_r$ and $E_0'$, becomes lower than $E_0$ in the region $E>E_r$.   
We determine 
the width of the effective energy window with Eq.~(\ref{eq:tdgamow}) by replacing $E_0$ by $E_0'$, that is by 
\begin{equation}
  \label{eq:tdgamow'}
  \Delta E_0'=\frac{4}{\sqrt{3}}(E_0'k_BT)^{1/2}.
\end{equation}
This definition of the width is different from the one chosen in~\cite{newton, rauscher2010}, but it shows  
clearly that the width is consistent with the Gamow window if the S-factor does not depend on the incident energy.  
We evaluate this condition numerically by using the $S$-factors determined experimentally for three selected
resonant reactions: $^{11}$B($p,\alpha$)$^8$Be, $^{10}$B($p,\alpha$)$^7$Be and $^3$H($d,n$)$^4$He, besides 
a non-resonant reaction $^2$H($d,n$)$^3$He. All four reactions are of importance in the application of the laser-induced 
nuclear reactions.

It is worth mentioning that the major contribution to the reaction rates comes from the 
vicinities of both Gamow energy and the resonant peak 
in resonant reactions is already known~\cite{nevins, ueda}, 
and it has been demonstrated  that the effective energy window in which the most thermonuclear reactions take place at a given 
temperature can differ significantly from Gamow peak~\cite{newton,rauscher2010}. 
Attention was focused on the ($p,\gamma$) reactions~\cite{newton}
and proton, $\alpha$ and neutron induced reactions~\cite{rauscher2010} on targets with 10 $\leq Z \leq$ 83 
at high temperatures (of the order of 1~GK~\cite{clayton}, i.e., $k_BT=$86 keV) which are relevant in the advanced burning stages of massive stars and in explosive stellar environments.   
Our aim in this paper is to attract attention on the fact that 
the effective energy window of the nuclear reaction driven by intense laser-pulse irradiation can deviate from the Gamow peak,
because 
the temperature region of the explosive stellar environment exactly matches the temperatures of 
the laser accelerated ions from a thin foil target by the TNSA mechanism and of 
the BOA mechanism~\cite{prl2013}. 
Whereas the plasma temperature of the laser-cluster fusion~\cite{woosk,  marina} is lower (30 keV at highest) than this criterion~\cite{kb-sc}.

\begin{figure}[h]
\begin{center}
\includegraphics[width=.37\textwidth, bb=80 80 382 322]{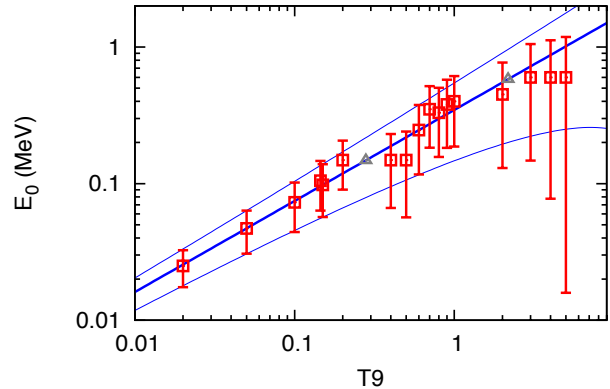}
\end{center}
\caption{\label{fig:te} (Color online) Most effective energy as a function of temperature for 
the reaction $^{11}$B($p,\alpha$)$^8$Be, where the abscissa
$T_9$ is the temperature in the unit of 10$^9$~K. 
The thick solid curve shows the relation~(\ref{eq:saddl}) and the effective energy range is the region between two thin curves.
the squares with error-bars are the most effective energy region at the corresponding temperature. The triangles 
show the positions of the resonance peaks.}
\end{figure}

\begin{figure}[h]
\begin{center}
\includegraphics[width=.35\textwidth, bb=80 80 382 352]{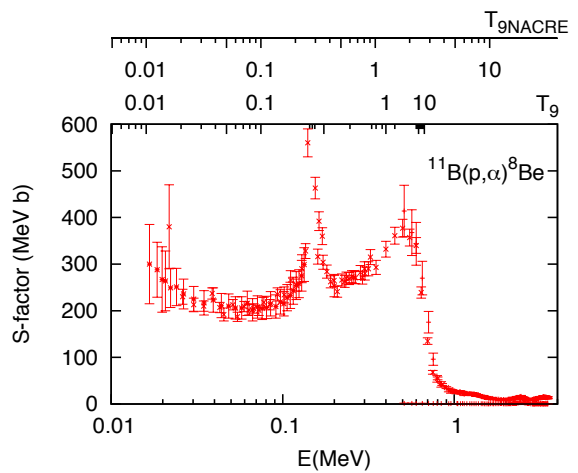}
\end{center}
\caption{\label{fig:11bp} (Color online) Correspondence between the plasma temperature and the most effective energy for the reaction 
$^{11}$B($p,\alpha$)$^8$Be. Experimental data of S-factor are retrieved from Ref.~\cite{be87}~(crosses),~\cite{an93}~(asterisks) 
and \cite{se65}~(bars).}
\end{figure}

We begin with 
the reaction $^{11}$B($p,\alpha$)$^8$Be, which has two low energy resonances 
at $E_r (\Gamma)=148$~keV (5.2~keV)
and at 581.3~keV  (300~keV).
Fig.~\ref{fig:te} is a plot of the most effective energies as a function of the plasma temperature for the reaction 
 $^{11}$B($p,\alpha$)$^8$Be.
The most effective energy in the presence of the low-energy resonances,  
evaluated from the condition~(\ref{eq:saddl-2}) (squares) is compared with the relation~(\ref{eq:saddl})~(solid line).
The most effective energy 
deviate clearly from the solid line at the resonant energies.
The $1/e$ width given by the Gamow peak approximation is shown by the region between two thin curves, while the width 
given by Eq.~(\ref{eq:tdgamow'}) is indicated as the error bars. The effective energy window deviates clearly from the one given by the Gamow 
peak approximation especially around the resonances.
Another point which should be remarked is that the effective energy window is widened as the temperature of the plasma rises, as it is clearly observed in the figure. 
With regard to the determination of the low energy cross section through the measurement of fusion yield in a laser-induced plasma, 
this means that the approximation of the effective energy by an energy is not adequate
especially in the high temperature region.
Fig.~\ref{fig:11bp} shows 
the correspondence between $T_9$ and $E$, which is derived from the relation~(\ref{eq:saddl-2}), together 
with the experimental data of the astrophysical S-factor for the reaction $^{11}$B($p,\alpha$)$^8$Be. 
For the sake of comparison, the $T_9$-axis from the relation~(\ref{eq:saddl}) is shown above the figure.  
Especially in the vicinity of the resonant peaks the change of the $T_9$ scale is evident.

\begin{figure}[h]
\begin{center}
\includegraphics[width=.35\textwidth, bb=80 80 382 372]{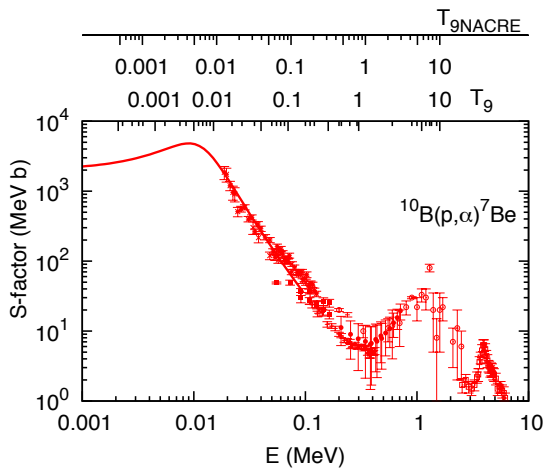}
\end{center}
\caption{\label{fig:10bp} (Color online) Same as Fig.\ref{fig:11bp} but for the reaction 
$^{10}$B($p,\alpha$)$^7$Be. Experimental data of S-factor has been
retrieved from NACRE compilation database, except the datasets shown by full circles~\cite{bu50} whose cross section data are 
retrieved from EXFOR database and are converted into the S-factor data. }
\end{figure}

Next, for the reaction $^{10}$B($p,\alpha$)$^7$Be
the experimental data of S-factor is shown in Fig.~\ref{fig:10bp}.
The $S$-factor increases as the incident energy decreases, 
this is interpreted as a part of a known s-wave resonance at the incident energy $E=$9.1~keV and with the width of $\Gamma=$16.~keV.
We include this resonance 
by using the Breit-Wigner formula. 
In Fig.~\ref{fig:10bp}   
the correspondence between $T_9$ and $E$  is shown, together 
with the experimental data of the astrophysical S-factor for the reaction $^{10}$B($p,\alpha$)$^7$Be.
Compared with $T_9$-axis from the relation~(\ref{eq:saddl}), which is shown above the figure,  
the change of the $T_9$ scale is evident around the s-wave resonance at $E_r=9.1$~keV. 
In the higher temperature region the change of the $T_9$ scale is less evident in contrast with the reaction $^{11}$B($p,\alpha$)$^8$Be.
It is because the $S$-factor in the $^{10}$B($p,\alpha$)$^7$Be is almost constant in higher temperature region. 
We note that the $S$-factor is shown in a logarithmic scale only for this reaction.   
Hereafter the figures of the effective energy range are not shown but the deviation from the Gamow peak approximation is 
seen in the effective energy range, as well.  

\begin{figure}[h]
\begin{center}
\includegraphics[width=.35\textwidth, bb=80 80 382 372]{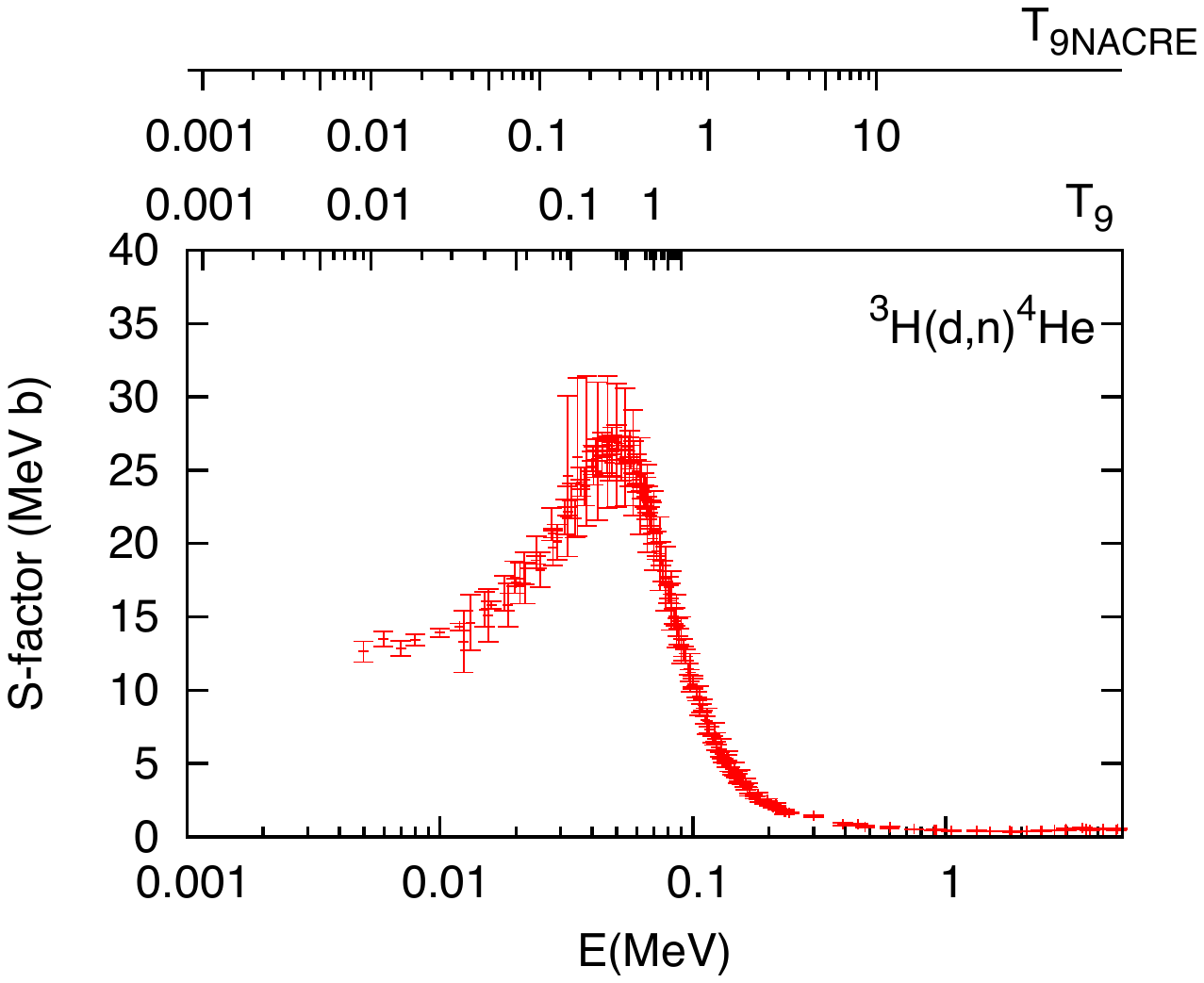}
\end{center}
\caption{\label{fig:td} (Color online) Same as Fig.\ref{fig:11bp} but for the reaction 
$^3$H($d,n$)$^4$He. Experimental data of S-factor has been retrieved from NACRE compilation database.}
\end{figure}

\begin{figure}[h]
\begin{center}
\includegraphics[width=.35\textwidth, bb=80 80 382 372]{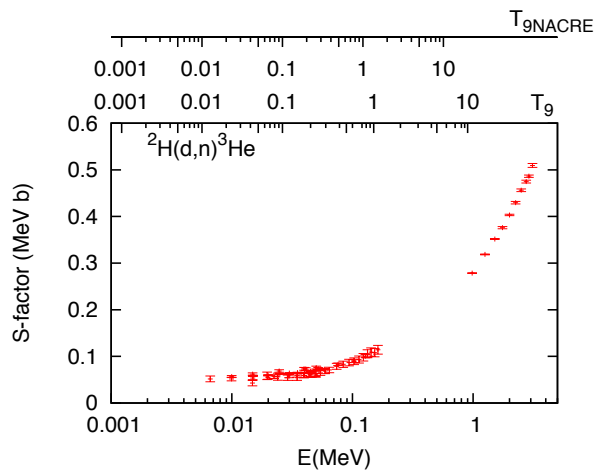}
\end{center}
\caption{\label{fig:dd} (Color online) Same as Fig.\ref{fig:11bp} but for the reaction 
$^2$H($d,n$)$^3$He. Experimental data of S-factor has been retrieved from NACRE compilation database.}
\end{figure}

The reaction $^3$H($d,n$)$^4$He has a resonance at $E_r(\Gamma)=50$~keV(177~keV).  
In Fig.~\ref{fig:td} the correspondence between $T_9$ and $E$ is shown
together 
with the experimental data of the astrophysical S-factor for this reaction.   
Compared with $T_9$-axis from the relation~(\ref{eq:saddl})
the change of the $T_9$ scale is clearly observed at about the low energy resonance.  
The last example is the reaction $^2$H($d,n$)$^3$He, which is non-resonant, but 
the $S$-factor of this reaction has slow energy dependence in the energy region above 50~keV, 
as is shown in Fig.~\ref{fig:dd}. 
The correspondence between $T_9$ and $E$  is shown together 
with the experimental data of the astrophysical S-factor for this reaction in the same figure.
Compared with $T_9$-axis from the relation~(\ref{eq:saddl}), which is shown above the figure,  
the $T_9$ scale shifts moderately toward higher energies, at the temperature $T_9$ higher than 0.3. 
This is attributed to 
the slow rise of the $S$-factor.  

In conclusion, we have evaluated the most effective energy region for charged particle induced reactions in plasma environment   
at a given plasma temperature. 
The correspondence between the plasma temperature and the most effective energy range is modified, 
especially where the astrophysical $S$-factor has a significant energy dependence.
We have shown this modifications for four selected reactions: 
$^{11}$B($p,\alpha$)$^8$Be, $^{10}$B($p,\alpha$)$^7$Be, $^3$H($d,n$)$^4$He and $^2$H($d,n$)$^3$He.
In the vicinity of the resonant peaks the change of the $T_9$ scale is remarkable.
In the presence of low-energy resonances, the resonances dominate the most effective energy.
The moderate change of the $T_9$ scale is observed also in the non-resonant reaction, in the energy region 
where the incident-energy dependence of the $S$-factor is significant. 
The suggested modification of the effective energy range is important not only in thermonuclear reactions at high temperature 
in advanced burning stages of massive stars and in explosive stellar environment but also in nuclear reactions driven by ultra-intense 
laser pulse irradiations.

\bibliography{memo.bib}
%



\end{document}